\documentclass{aa}

\usepackage{natbib}
\usepackage{epsfig}
\usepackage{txfonts}

\begin{document} 

\newcommand{\chisq}{\mbox{$\chi^2$}}
\newcommand{\es}{erg s$^{-1}$}                          
\newcommand{\ecms}{erg cm$^{-2}$ s$^{-1}$}              
\newcommand{\halpha}{H$\alpha$}                   
\newcommand{\hbeta}{H$\beta$}
\newcommand{\kms}{km~s$^{-1}$}       
\newcommand{\cmthree}{cm$^{-3}$}
\newcommand{\msun}{M$_{\odot}$} 
\newcommand{\xmm}{XMM-\emph{Newton}} 
\newcommand{\chandra}{\emph{Chandra}} 
\newcommand{\spitzer}{\emph{Spitzer}} 
\newcommand{\droxo}{{\sc Droxo}}
\newcommand{\nh}{\mbox{$N({\rm H})$}}
\newcommand{\rhoph}{$\rho$~Oph}
\newcommand{\eltn}{Elias 29}
\newcommand{\eltf}{Elias 24}
\newcommand{\hlt}{HL~Tau}

\title{Results from {\sc Droxo}. I. The variability of fluorescent Fe 6.4 keV emission in the
young star Elias~29}

\subtitle{High-energy electrons in the star's accretion tubes?}

\author{G. Giardino\inst{1} \and F. Favata\inst{2} \and I. Pillitteri\inst{3} \and
  E. Flaccomio\inst{3} \and G. Micela\inst{3}  \and S. Sciortino\inst{3}}

\institute{Astrophysics Division -- Research and Science Support
  Department of ESA, ESTEC, 
  Postbus 299, NL-2200 AG Noordwijk, The Netherlands
\and
ESA -- Planning and Community Coordination Office, Science Programme, 8-10 rue Mario Nikis,  F-75738 Paris Cedex 15, France
\and
INAF -- Osservatorio Astronomico di Palermo, 
Piazza del Parlamento 1, I-90134 Palermo, Italy 
}

\offprints{G. Giardino}

\date{Received date / Accepted date}

\abstract 
  {} {We study the variability of the Fe 6.4 KeV emission line from
  the Class I young stellar object Elias 29 in the $\rho$ Oph cloud.}
  {We analysed the data from Elias 29 collected by \xmm\ during a
  nine-day, nearly continuous observation of the $\rho$ Oph
  star-forming region (the Deep Rho-Oph X-ray Observation, named {{\sc
  Droxo}}). The data were subdivided into six homogeneous time
  intervals, and the six resulting spectra were individually analysed}
  {We detect significant variability in the equivalent width of the Fe
  6.4 keV emission line from Elias 29. The 6.4~keV line is absent
  during the first time interval of observation and appears at its
  maximum strength during the second time interval (90 ks after Elias
  29 undergoes a strong flare). The X-ray thermal emission is
  unchanged between the two observation segments, while line
  variability is present at a 99.9\% confidence level. Given the
  significant line variability in the absence of variations in the
  X-ray ionising continuum and the weakness of the photoionising
  continuum from the star's thermal X-ray emission, we suggest that
  the fluorescence may be induced by collisional ionisation from an
  (unseen) population of non-thermal electrons. We speculate on the
  possibility that the electrons are accelerated in a reconnection
  event of a magnetically confined accretion loop, connecting the
  young star to its circumstellar disk.} {}


\maketitle

\section{Introduction}
\label{sec:intro}

The X-ray emission from young stellar objects (YSOs) at CCD-resolution
is usually modelled as thermal emission from a hot plasma in coronal
equilibrium, with higher characteristic temperatures than observed in
older and less active stars.  An interesting deviation from a pure
thermal X-ray spectrum is the presence of fluorescent emission from
neutral (or weakly ionised) Fe as shown by the presence of the 6.4 keV
line. This was first detected by \citet{ikt01} in the X-ray emission
of the YSO YLW16A in $\rho$-Oph, during a large flare: in addition to
the Fe\,{\sc xxv} complex at 6.7 keV, a 6.4 keV emission line was
clearly visibe. 
Such fluorescence line is produced when energetic
X-rays photoionise cold material close to the X-ray source, and it is
therefore a useful diagnostic tool of the geometry of the X-ray
emitting source and its surroundings.  

Since 2001, detections of the Fe K fluorescent emission line at
6.4-keV in the spectra of YSOs have been reported by a number of authors.
\citet{tfg+05} has identified seven sources with an excess emission at 6.4
keV among 127 observations of YSOs within the COUP observation of
Orion; \citet{fms+05} report 6.4 keV fluorescent emission in Elias
29 in $\rho$-Oph both during quiescent and flaring emission, unlike
all other reported detection of Fe fluorescent emission in YSOs that
were made during intense flaring; \citet{gfm+07} have detected Fe 6.4 keV
emission from a low-mass young star in Serpens, during an intense,
long-duration flare. Recently, \cite{sc2007} have reported
intense Fe fluorescent emission in the spectrum of V 1486 Ori during a
strong flare, when the plasma reached a temperature in excess of 10
keV.

The 6.4 keV fluorescent line has been detected in different classes of
X-ray emitters: X-ray binaries, active galactic nuclei (AGNs), massive
stars, supernova remnants, and the Sun itself during flares. In the
case of the Sun, the fluorescing material is the solar photosphere, in
the YSOs, however, indications are that the material in the
circumstellar disk and its related accretion structures could be
responsible for the fluorescence. The typical equivalent width of the
6.4 keV emission line in the studies mentioned above is of the order
of 150~eV and is too large to be explained with fluorescent emission
in the stellar photosphere or in diffuse circumstellar material
(e.g. \citealp{tfg+05}; \citealp{fms+05}).  This scenario implies that
the disk is ``bathed'' in high-energy X-rays emitted by the star, with
significant astrophysical implications; for instance, X-rays, in
addition to cosmic rays, would play an important role in photoionising
the circumstellar material around young star and thus in coupling the
gas to the ambient magnetic field (as suggested by
e.g. \citealp{gfm00}). \citet{cbt+02} suggest that the ``hot''
component they observe in the disk, in the infrared, is heated by the
stellar high-energy emission.

\begin{figure*}
  \begin{center} 
	\leavevmode 
	\epsfig{file=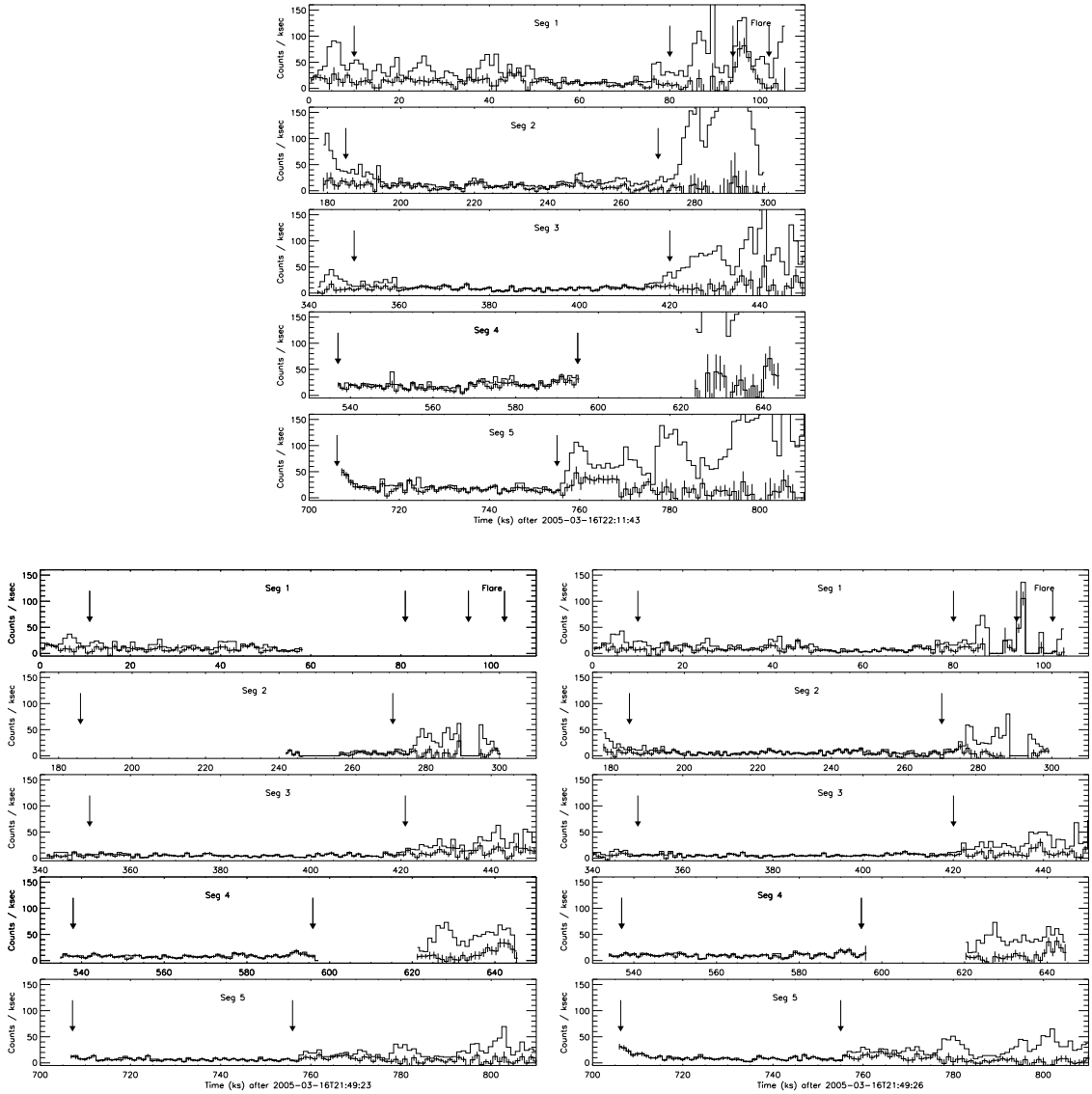, width=17.0cm, bbllx=0,bblly=190,bburx=842,bbury=1000, angle=270, clip=}
        
\caption{The light curve of Elias 29 over the nine days of the \droxo\
observation from PN ({\em top}), MOS1 ({\em bottom-left}), and MOS2
({\em bottom-right)}. The line with error bars gives the
background-subtracted light curve of the source, while the thinner
line without error bars gives the total counts (source plus
background). The 6 time intervals that we selected for the spectral
analysis, on the basis of the PN data, are indicated.}

    \label{fig:lc}
  \end{center}
\end{figure*}

Observations of the time variability of the Fe 6.4 keV emission in
YSOs would provide useful constraints on the geometry and sizes of the
star-accretion disk system and of other circumstellar structures such
as funnel flows, jets, or wind columns. We present here the results of
a time-resolved spectral study of the X-ray emission of Elias 29,
during $\sim 9$ days of nearly continuous observation by \xmm\ in the
context of the ultra-deep observation of $\rho$-Oph, named \droxo\
(from Deep Rho-Oph X-ray observation -- \citealp{psf+07}).  We
investigated the presence of variations in its strong Fe 6.4 keV
emission over the 9-day time scale covered by the observations.

This paper is structured as follows. After a summary, below, of the
properties of Elias 29, the observations and data analysis are briefly
presented in Sect.\,\ref{sec:obs}. Results are summarised in
Sect.\,\ref{sec:res}; the simulations carried out to assess the
reliability of the line detections and the significance of its
variability are described in Sect.\,\ref{sec:sim}. The results are
discussed in Sect.\,\ref{sec:disc}.

\subsection{Elias 29 (GY214)}

Elias 29 (16:27:09.4, $-$24:37:18.9) with a bolometric luminosity $L =
26-27.5~L_{\sun}$ (\citealp{bak+01}; \citealp{nts06}) is the most
luminous Class I YSO in the $\rho$-Oph cloud. \cite{mhc98} used the
luminosity in the Br$\gamma$ line to determine the object's accretion
luminosity at $L_{\rm acc}$ = $15-18 L_{\sun}$, which makes it the
source with the highest accretion luminosity in their sample. More
recently, \citet{nts06} used the luminosity of the hydrogen
recombination lines to derive an accretion
luminosity of 28.8~$L_{\sun}$.

Using millimeter interferometric observations, \citet{bhc+02} resolved
the emission from the disk and the envelope surrounding Elias 29,
showing that the disk is in a relatively face-on orientation ($i <
60^{\circ}$), which explains many of the remarkable observational
features of this source, such as its flat spectral energy
distribution, its brightness in the near-infrared, the extended
components found in speckle interferometry observations, and its
high-velocity molecular outflow.
Their best-fitting disk model has an inner radius
of 0.01 AU, outer radius of 500 AU, and a mass
$M = 0.012 M_{\sun}$. 

Elias 29 was previously observed in X-rays with ASCA, \chandra\, and
\xmm.  In the \chandra\ observation (\citealp{ikt01}), the source
quiescent phase is characterised by a temperature of 4.3~keV and
luminosity of $2.0 \times 10^{30}$~\es, fully consistent with the
values derived from the subsequent \xmm\ observations by
\citet{ogm05}: $kT = (3.6-5.1)$ keV, $N({\rm H}) = (4.4-5.3) \times
10^{22}$ cm$^{-2}$, $Z=(0.8-1.3)~Z_{\sun}$, and $L_{\rm X} = 2.8 \times
10^{30}$ \es.  The source was seen flaring during one of the ASCA
observations (\citealp{kkt+97}) and during the \chandra\
observation. The two flares had similar intensity and duration
with an $e$-folding time of $\sim 10$~ks (\citealp{tik+00};
\citealp{ikt01}).

\section{Observations and data analysis}
\label{sec:obs}

The {\sc Droxo} program is a nominal 500 ks observation of the
$\rho$-Ophiuchi star-forming region performed by the EPIC camera on
board the \xmm\ satellite. The observation was performed over 9.4
days, starting 8 March 2005 (orbits 0961--0965). Details of the
observations and data reduction procedure are given in \citet{psf+07}.
The preliminary data reduction was done with SAS software version 6.5
in order to obtain lists of photon events calibrated both in energy
and astrometry for the three instruments, MOS1, MOS2, and PN, for
each orbit. The data were filtered in the energy band 0.3--10. keV,
and only events with {\sc pattern} $<4$ and {\sc flag} $=0$ were
retained for the spectral analysis. The spectral analysis was
performed using the \textsc{xspec} package V11.2, after rebinning the
spectra to a minimum of 20 source counts per (variable width) spectral
bin.


\section{Results}
\label{sec:res}

Figure \ref{fig:lc} shows the light curve of Elias 29 from the
three instruments, PN, MOS1, and MOS2, and the six time intervals
that we selected for our spectral analysis. On the basis of the
PN data, we selected 5 time intervals with low background (hereafter
``seg1'' to ``seg5''), plus one time interval covering the strong
flare at about 94 ks from the beginning of the observation. During
``seg1'', the MOS1 camera was very likely hit by a micro-meteorite
(which compromised one of the chips) and the instrument was switched
off for the rest of this segment of the observation, during the flare
and part of ``seg2''. MOS2 data are available for all the segments of
observation, but are insufficient during the flare, thus, we did not
include the ``flare'' time interval in the joint spectral analysis of
PN and MOS data. In addition, we only used PN and MOS2 data
for ``seg1'' and ``seg2''.

The spectra were initially modelled by an absorbed one-temperature
plasma model. The results of the six spectral fits for the PN
data alone are reported in Table \ref{tab:psfit} and the results of
the (five) joint fits to the PN, MOS1, and MOS2 data in
Table \ref{tab:psfit_mos}. For each time-interval, the values of the
spectral parameters derived from the simultaneous fitting are very
similar to the values derived from the PN data alone. The error bars
in the best-fit parameters improve marginally, but $\chi^2$ values
worsen, likely because of calibration uncertainties.

The spectra and the fits for ``seg1'' and ``seg2'' are shown in
Fig.\,\ref{fig:psfit} and the spectrum for the flare in
Fig.\,\ref{fig:psflare}.  The average values of the source's spectral
parameters while quiescent ($N({\rm H}) = 6.8 \times 10^{22}$
cm$^{-2}$, $kT = 3.7$ keV, and $Z=0.8~Z_{\sun}$) are very similar to
the values derived from previous observations, so is its quiescent
luminosity $L_{\rm X} \sim 10^{30}$~\es, with no evidence of long-term
variability.

The X-ray flare is similar to other events previously observed from
this source. During the flare, the source counts first increased
impulsively by a factor of $\sim 8$ and then decreased exponentially
with a decay time of $\sim 6$~ks.  The source spectrum during the
flare is shown in Fig.\,\ref{fig:psflare}, together with the spectral
fit. As presented in Table\,\ref{tab:psfit}, the fitted value of the
plasma temperature does not appear to change significantly during the
flare, possibly due to the stringent processing criteria
applied to the data that resulted in the events of the flare peak
being discarded.

From Tables\,\ref{tab:psfit} and \ref{tab:psfit_mos}, it is apparent
that the best-fit values of $N({\rm H})$, $kT$, and $Z$ for the
different time intervals do not show significant variations, since
they are all consistent with each other within $2\sigma$. This is also
apparent by comparing the PN and MOS2 spectra from ``seg1'' and
``seg2'' in Fig.\,\ref{fig:psfit}, where the overall spectral shape and
amplitude are very similar during the two time intervals (in both
instruments). On closer inspection, however, a significant difference
between the two time intervals becomes apparent: during time interval
``seg2'', a visible excess of emission around 6.4 keV, the energy of
the Fe fluorescent line, is present both in the PN and MOS2
data.

To quantify this excess and monitor its variation, we repeated the
spectral fits of the spectra with an absorbed 1T plasma model and
an additional Gaussian line component at 6.4 keV. The position and
width (10 eV\footnote{The iron K$\alpha$ fluorescence line
consists of two components with an energy separation of $\sim 10$ eV:
K$\alpha_1$ and K$\alpha_2$ at 6.404 and 6.391 keV, respectively, for
Fe\,{\sc i}}) of this component were constrained during the fit,
while its normalisation was left free to vary. The other parameters of
the absorbed 1T plasma model (absorbing column density, temperature,
and normalisation of the thermal spectrum) were also free parameters
in the fit. The spectra and spectral fits to the six PN spectra in the
energy range $\Delta E = 4 - 8$~keV are shown in
Fig.\,\ref{fig:ps_line}. The strong excess emission at 6.4 keV in the
spectra of ``seg2'' is well accounted for by the fitted line, while it
is clear that no such additional line at 6.4 keV is needed to fit the
data from ``seg1''.  The four other spectra are somewhat in between
these two extremes in regard to an excess of emission at 6.4 keV.

The results from the spectral fits  to the PN data with the
additional line at 6.4 keV are summarised in Table
\ref{tab:psfit_line}.  As can be seen by comparing Tables
\ref{tab:psfit} and \ref{tab:psfit_line}, the fitted values for
absorbing column density, temperature, metallicity, and normalisation
of the thermal spectrum are not affected by the addition of the 6.4
keV-line component. The null-hypothesis probability of all the fits
improves, except for ``seg1'', for which the null-hypothesis
probability decreases marginally.  Indeed, an excess of emission at
6.4 keV does not appear to be present at all in this spectrum, so that
there is no reason to add an extra line component at 6.4 keV. We have
also added this component to the spectral fit of ``seg1'' to make the
fits homogeneous and directly comparable.  The equivalent width of the
fitted line appears to vary significantly from the 13 eV of ``seg1''
(where the excess emission at 6.4 keV does not seem to be present) to
the 249 eV of ``seg2''.

Table\,\ref{tab:psfit_line_mos} summarises the results of the
simultaneous spectral fitting of PN and MOS data with an absorbed 1T
plasma model and an additional Gaussian line component at 6.4 keV: the
results obtained from this joint spectral analysis are again fully
consistent with the results from the PN data alone (compare
Table\,\ref{tab:psfit_line} and \ref{tab:psfit_line_mos}). The strong
variation in the intensity of the excess emission at 6.4 keV between
``seg1'' and ``seg2'' is confirmed; the best-fit line equivalent width
varies from 10 eV in ``seg1'' (where the line also appears to be
absent in the MOS2 data) to 194 eV in ``seg2'' (where the excess of
emission is also strong in the MOS2 data -- see
Fig.\,\ref{fig:psfit}).  For ``seg3'' and ``seg4'', the equivalent
width of the emission at 6.4 keV derived from the joint fits differs
by $\sim $50\% from the value derived from the PN data alone, but
best-fit values of the line's flux are consistent within the two
data-sets; for ``seg5'', the equivalent width of the emission at 6.4
keV derived from the two spectral analysis are very similar.

As described in Sect.\,\ref{sec:sim}, we have performed a set of Monte
Carlo simulations of the PN spectra to assess whether the
line at 6.4 keV seen in the spectra from ``seg2'' to ``seg5'' is real
rather than the result of statistical fluctuations in the data. The
results of these simulations are summarised in the last column of
Table\,\ref{tab:psfit_line}, which gives the probability that a line
at 6.4 keV, with equivalent width equal to, or greater than, the
best-fit value, could be the result of random fluctuations for each
segment. This probability is high for ``seg1'' and the ``flare''
segment, for which the data are consistent with the absence of a line
in the spectrum.  Conversely, the probability that the excess of
emission in ``seg2'' and ``seg5'' is due to noise fluctuations is low
(0.1\% and 0.2\%, respectively).  For time intervals ``seg3'' and
``seg4'' the situation is less clear, since with a probability of 4.2\% and
3.5\%, respectively, that the excess at 6.4 keV could be due to random
fluctuations, the evidence for intrinsic emission is less compelling.
Nevertheless an analysis of the spectrum of the source integrated over
the time intervals ``seg3'', ``seg4'', and ``seg5'' shows that the
evidence for the presence of the line is very strong, with a
probability that the excess at 6.4~keV is due to random fluctuation of
less than one in a thousand (as summarised in the last line of
Table\,\ref{tab:psfit_line}).

\begin{figure*}
  \begin{center} \leavevmode 

	\epsfig{file=7899fg02.ps, height=8.0cm, bbllx=80, bblly=43, bburx=535, bbury=700, angle=270, clip=}
        \epsfig{file=7899fg03.ps, height=8.0cm, bbllx=80, bblly=43, bburx=535, bbury=700, angle=270, clip=}
	\epsfig{file=7899fg04.ps, height=8.0cm, bbllx=80, bblly=43, bburx=550, bbury=700, angle=270, clip=}
	\epsfig{file=7899fg05.ps, height=8.0cm, bbllx=80, bblly=43, bburx=550, bbury=700, angle=270, clip=} 

\caption{Elias 29 spectra from PN ({\em top}) and MOS2 ({\em bottom})
with spectral fits, during ``seg1'' (before the flare, {\em left}) and
``seg2'' (after the flare, {\em right}). The spectra (from the two
time intervals) are very similar in overall shape and amplitude, and
the fits (with an absorbed 1T plasma model) result in very similar
parameters. During ``seg2'', however, a significant excess of emission
at 6.4 keV is present, which is not visible in the data from ``seg1''. All
the spectra are rebinned to a minimum of 20 source counts per
(variable width) spectral bin.}

    \label{fig:psfit}
  \end{center}
\end{figure*}

\begin{figure}
  \begin{center} \leavevmode 
        \epsfig{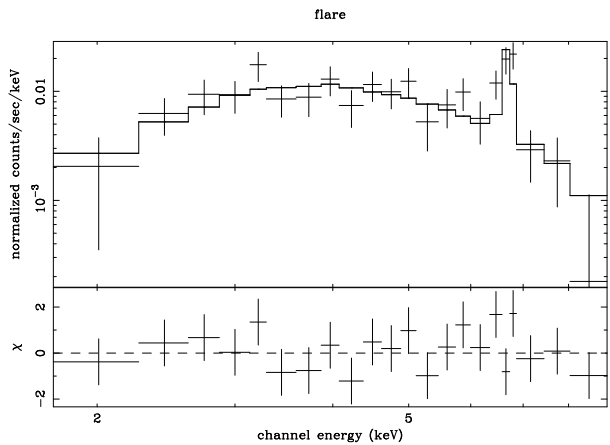}

\caption{The spectrum of Elias 29 (PN data) during the strong flare
between ``seg1'' and ``seg2'', together with the best-fit
one-temperature model.}
    
\label{fig:psflare}
  \end{center}
\end{figure}

The simulations described in Sect.\,\ref{sec:sim}  also show
that the variation in the line's equivalent width seen between ``seg1'' (where
the line appear to be absent or extremely weak) and ``seg2'' are very
likely intrinsic (with a probability of 99.9\%).

\begin{table*}[h]
  \begin{center}

  \caption{Best-fit parameters from the fits to the spectra of Elias
  29, during 6 different time intervals (PN data only), using an absorbed 1T
  thermal model. Reduced $\chi^2$ and null-hypothesis
  probability ($P$) for the fits are also given. }

    \leavevmode
     \begin{tabular}{l|ccccccc}
Time interval &  $N({\rm H})$ & $kT$ & $Z$ & $E\!M$ & $\chi^2$/d.o.f. & $P$ & $F_{\rm X}$\\
\hline
~ & $N_{22}$ & keV &  $Z_{\odot}$ & $E\!M_{53}$ & ~ & ~ & $F_{-13}$ \\
\hline
seg1 & 5.7 $\pm$ 0.7 & 4.6 $\pm$ 0.9 & 1.0 $\pm$ 0.3 & 1.4 $\pm$ 0.5 & 1.0 & 0.6 & 1.8\\
flare & 9.5 $\pm$ 2.0 & 4.1 $\pm$ 1.5 & 0.9 $\pm$ 0.3 & 8.3 $\pm$ 6.3 & 0.9 & 0.5 & 7.5\\
seg2 & 8.1 $\pm$ 1.3 & 3.1 $\pm$ 0.8 & 1.0 $\pm$ 0.3 & 1.4 $\pm$ 0.9 & 1.5 & 0.0 & 1.1\\
seg3 & 6.6 $\pm$ 0.8 & 3.4 $\pm$ 0.7 & 0.7 $\pm$ 0.2 & 1.5 $\pm$ 0.8 & 1.0 & 0.5 & 1.3\\
seg4 & 7.7 $\pm$ 0.6 & 3.7 $\pm$ 0.5 & 0.7 $\pm$ 0.1 & 3.1 $\pm$ 0.9 & 1.4 & 0.0 & 2.7\\
seg5 & 4.9 $\pm$ 0.4 & 4.7 $\pm$ 0.8 & 0.6 $\pm$ 0.1 & 2.0 $\pm$ 0.6 & 1.3 & 0.1 & 2.5\\
\hline
    \end{tabular}
    \label{tab:psfit}
  \end{center}

Units are $N_{22} = 10^{22}~{\rm cm^{-2}}$, $E\!M_{53} =
10^{53}$~cm$^{-3}$, and $F_{-13}= 10^{-13}$~\ecms. The spectral fits
were carried out in the energy range 1.0--9.0~keV. Flux values refer
to energy range 1.0-7.5 keV. The error bars are at 1$\sigma$.

\end{table*}

\begin{table*}[h]
  \begin{center}

  \caption{Best-fit parameters from the joint fits to PN, MOS1
  and, MOS2 spectra of Elias 29, during 5 different time intervals,
  using an absorbed 1T thermal model.}

    \leavevmode
     \begin{tabular}{l|ccccccc}
Time interval &  $N({\rm H})$ & $kT$ & $Z$ & $E\!M$ & $\chi^2$/d.o.f. & $P$ & $F_{\rm X}$\\
\hline
~ & $N_{22}$ & keV &  $Z_{\odot}$ & $E\!M_{53}$ & ~ & ~ & $F_{-13}$ \\
\hline
seg1\dag & 5.5 $\pm$ 0.5 & 4.4 $\pm$ 0.7 & 1.0 $\pm$ 0.2 & 1.7 $\pm$ 0.5 & 1.1 & 0.1 & 2.1\\
seg2\dag & 7.4 $\pm$ 0.8 & 3.8 $\pm$ 0.7 & 0.8 $\pm$ 0.2 & 1.5 $\pm$ 0.6 & 1.6 & 0.0 & 1.3\\
seg3 & 6.6 $\pm$ 0.5 & 3.5 $\pm$ 0.5 & 0.7 $\pm$ 0.1 & 2.0 $\pm$ 0.5 & 1.1 & 0.2 & 1.7\\
seg4 & 7.5 $\pm$ 0.4 & 3.7 $\pm$ 0.4 & 0.7 $\pm$ 0.1 & 3.5 $\pm$ 0.8 & 1.5 & 0.0 & 3.0\\
seg5 & 5.7 $\pm$ 0.4 & 4.0 $\pm$ 0.5 & 0.5 $\pm$ 0.1 & 2.8 $\pm$ 0.7 & 1.6 & 0.0 & 2.3\\
\hline
	\end{tabular}
\label{tab:psfit_mos}
  \end{center}

\dag MOS1 data (mostly) unavailable for this time interval so only PN and
MOS2 data were used.    
\end{table*}

\begin{table*}[h]
  \begin{center}

  \caption{Best-fit parameters from the fits to the spectra of Elias
  29, during 6 different time intervals (PN data only), using an
  absorbed 1T thermal model with an additional line at 6.4 keV (Fe
  fluorescent line). For the line, the total flux in the line ($f_{\rm
  6.4 keV}$ in units of $f_{-6} = 10^{-6}$ photons cm$^{-2}$ s$^{-1}$)
  and the equivalent width ($W_{\rm 6.4~keV}$) are given.}

    \leavevmode
     \begin{tabular}{l|ccccccccc|c}
Time interval &  $N({\rm H})$ & $kT$ & $Z$ & $E\!M$ & $f_{\rm 6.4 keV}$ & $W_{\rm 6.4~keV}$ & $\chi^2$/d.o.f. & $P$ & $F_{\rm X}$ & $R_{\rm 6.4 keV}$\ddag \\
\hline
~ & $N_{22}$ & keV &  $Z_{\odot}$ & $E\!M_{53}$ & $f_{-6}$ & eV & ~ & ~ & $F_{-13}$ & \% \\
\hline

seg1 & 5.5 $\pm$ 0.7 & 4.8 $\pm$ 1.0 & 1.1 $\pm$ 0.3 & 1.3 $\pm$ 0.5 &
0.1 $\pm$ 0.7 & 13.0 & 1.0 & 0.5 & 1.8 & 45\\
flare & 9.4 $\pm$ 2.0 & 3.9 $\pm$ 1.4 & 0.9 $\pm$ 0.3 & 8.3 $\pm$ 6.2
& 2.0 $\pm$ 5.9 & 60.2  & 0.9 & 0.6 & 7.5 & 23\\
seg2 & 8.2 $\pm$ 1.3 & 2.8 $\pm$ 0.6 & 1.1 $\pm$ 0.3 & 1.4 $\pm$ 0.8 &
1.1 $\pm$ 0.6 & 249.0  & 1.3 & 0.1 & 1.1 & 0.1\\
seg3 & 6.6 $\pm$ 0.8 & 3.3 $\pm$ 0.6 & 0.7 $\pm$ 0.2 & 1.6 $\pm$ 0.8 &
0.5 $\pm$ 0.5 & 108.0  & 0.9 & 0.6 & 1.3 & 4.2\\
seg4 & 7.6 $\pm$ 0.6 & 3.6 $\pm$ 0.4 & 0.7 $\pm$ 0.1 & 3.2 $\pm$ 0.9 &
0.7 $\pm$ 0.7 & 67.8  & 1.3 & 0.1 & 2.7 & 3.5\\
seg5 & 4.9 $\pm$ 0.4 & 4.5 $\pm$ 0.8 & 0.6 $\pm$ 0.1 & 2.1 $\pm$ 0.6 &
1.1 $\pm$ 0.8 & 162.0  & 1.2 & 0.2 & 2.5 & 0.2\\
\hline
seg3$+$seg4$+$seg5$\diamondsuit$ & 6.4 $\pm$ 0.3 & 3.9 $\pm$ 0.3 & 0.6 $\pm$ 0.1 &
2.2 $\pm$ 0.4 & 1.0 $\pm$ 0.3 & 147.0  & 1.3 & 0.0 & 2.1 & 0.0\\
\hline
    \end{tabular}
    \label{tab:psfit_line}
  \end{center}

\ddag Probability that a line with a best-fit
value equal or greater than $W_{\rm 6.4~keV}$ could be due to a random
fluctuation in the data, as estimated via Monte Carlo simulations (see
Sect. 3 and 4).

$\diamondsuit$ Results for the fit to the source's spectrum integrated over
time intervals ``seg3'', ``seg4'', and ``seg5''

\end{table*}

\begin{table*}[h]
  \begin{center}

  \caption{Best-fit parameters from joint fits to PN, MOS1 and MOS2
  spectra of Elias 29, during 5 different time intervals, using an
  absorbed 1T thermal model with an additional line at 6.4 keV (Fe
  fluorescent line).}

    \leavevmode
     \begin{tabular}{l|ccccccccc}
Time interval &  $N({\rm H})$ & $kT$ & $Z$ & $E\!M$ & $f_{\rm 6.4 keV}$ & $W_{\rm 6.4~keV}$ & $\chi^2$/d.o.f. & $P$ & $F_{\rm X}$ \\
\hline
~ & $N_{22}$ & keV &  $Z_{\odot}$ & $E\!M_{53}$ & $f_{-6}$ & eV & ~ & ~ & $F_{-13}$ \\
\hline

seg1\dag & 5.5 $\pm$ 0.5 & 4.3 $\pm$ 0.8 & 1.0 $\pm$ 0.2 & 1.7 $\pm$ 0.5 & 0.1 $\pm$ 0.7 & 9.7  & 1.2 & 0.1 & 2.1\\
seg2\dag & 7.5 $\pm$ 0.8 & 3.5 $\pm$ 0.7 & 0.9 $\pm$ 0.2 & 1.6 $\pm$ 0.6 & 1.1 $\pm$ 0.7 & 194.0  & 1.4 & 0.0 & 1.3\\
seg3 & 6.7 $\pm$ 0.6 & 3.3 $\pm$ 0.5 & 0.7 $\pm$ 0.2 & 2.1 $\pm$ 0.7 & 0.4 $\pm$ 0.6 & 68.7  & 1.1 & 0.3 & 1.7\\
seg4 & 7.5 $\pm$ 0.4 & 3.5 $\pm$ 0.3 & 0.7 $\pm$ 0.1 & 3.6 $\pm$ 0.8 & 1.2 $\pm$ 0.7 & 118.0  & 1.4 & 0.0 & 3.0\\
seg5 & 5.8 $\pm$ 0.4 & 3.7 $\pm$ 0.5 & 0.5 $\pm$ 0.1 & 2.9 $\pm$ 0.8 & 1.2 $\pm$ 0.7 & 152.0 & 1.5 & 0.0 & 2.3\\

\hline
	\end{tabular}
\label{tab:psfit_line_mos}
  \end{center}

\dag MOS1 data (mostly) unavailable for this time interval so only PN and
MOS2 data were used.    
\end{table*}

\begin{figure*}
  \begin{center} \leavevmode 
        \epsfig{file=7899fg07.ps, height=8.0cm, angle=270}
        \epsfig{file=7899fg08.ps, height=8.0cm, angle=270}
        \epsfig{file=7899fg09.ps, height=8.0cm, angle=270}
        \epsfig{file=7899fg10.ps, height=8.0cm, angle=270}
	\epsfig{file=7899fg11.ps, height=8.0cm, angle=270}	
        \epsfig{file=7899fg12.ps, height=8.0cm, angle=270}

\caption{Spectra (PN) and spectral fits of Elias 29 during 6 different
time intervals, between 4 and 8 keV. From left to right, top to
bottom, spectra are from time intervals ``seg1'', ``flare'', ``seg2'',
``seg3'', ``seg4'', and ``seg5'' (these intervals are indicated in
Fig. 1). Fits were performed using an absorbed 1T thermal model with
the addition of a line at 6.4 keV (indicated in the plots by a
dash-dotted line).  Note the strong excess of emission at 6.4 keV in
the spectra for ``seg2'', which does not appear to be present in the
spectrum from the first time interval.}

\label{fig:ps_line}
\end{center}
\end{figure*}

\begin{figure}
  \begin{center} \leavevmode
        \epsfig{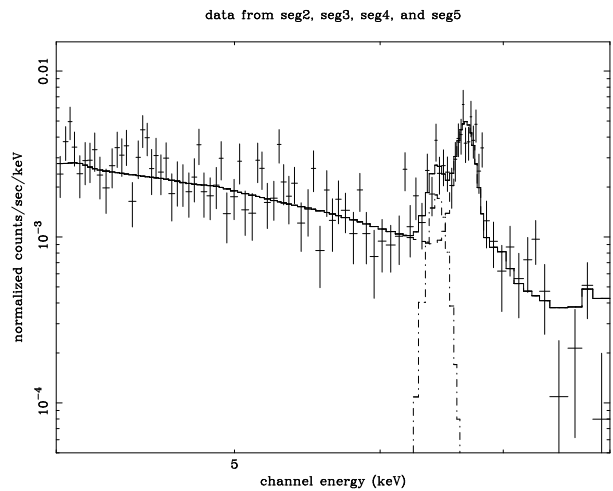}
\caption{The spectrum (PN data) of Elias 29, between 4 and 8 keV,
integrated over the four time intervals of quiescent emission where an
excess at 6.4 keV appears to be present. With the improved statistic of
this long time span, the presence of 6.4 keV Fe fluorescent emission
is very clear.}  

\label{fig:psint} 
\end{center}
\end{figure}

To further constrain the properties of the Fe 6.4~keV emission from
Elias 29, we derived the spectrum of the source integrated over the
quiescent time intervals where the excess at 6.4 keV appears to be
present (although with varying significance), i.e. intervals ``seg2'',
``seg3'', ``seg4'', and ``seg5'' (PN data only). The summed
spectrum between 4 and 8 keV is shown in Fig.\,\ref{fig:psint}, where
the emission line at 6.4 keV is very clear thanks to the the
improved statistic. A fit with an absorbed 1T model to this spectrum
yields $N({\rm H}) = 6.7 \times 10^{22}$ cm$^{-2}$, $kT = 3.7$ keV and
$Z=0.7~Z_{\sun}$ with a null hypothesis probability $P = 3.5 \times
10^{-5}$. The spectral parameters are very similar to the average
values derived above from the fits to the individual five quiescent
time intervals, although the fit probability is rather low. With the
addition of a line at 6.4 keV, the fit probability increases
substantially to $P=2.1 \times 10^{-3}$ (and the values of the
parameters of the 1T model hardly change). In this case we left the
position of the Gaussian line free to vary during the fit, together
with its normalisation. The fitted line equivalent width is $W_{\rm
6.4~keV} = 143$ eV and its position is $6.44 \pm 0.03$, fully
consistent with being Fe K fluorescent emission. Indeed,
although the reliability is limited (the quoted error is at
1$\sigma$), the line peak energy of 6.44 keV would suggest that the
fluorescing Fe is Neon-like: Fe\,{\sc xvii} or so. The Fe K
fluorescent emission line energy is, in fact, a slowly increasing
function of ionisation state, rising from 6.40 keV in Fe\,{\sc i} to
6.45 keV in Fe\,{\sc xvii} (\citealp{house69}; \citealp{gf91}).

\section{Simulations}
\label{sec:sim}

\begin{figure}
  \begin{center} \leavevmode 

        \epsfig{file=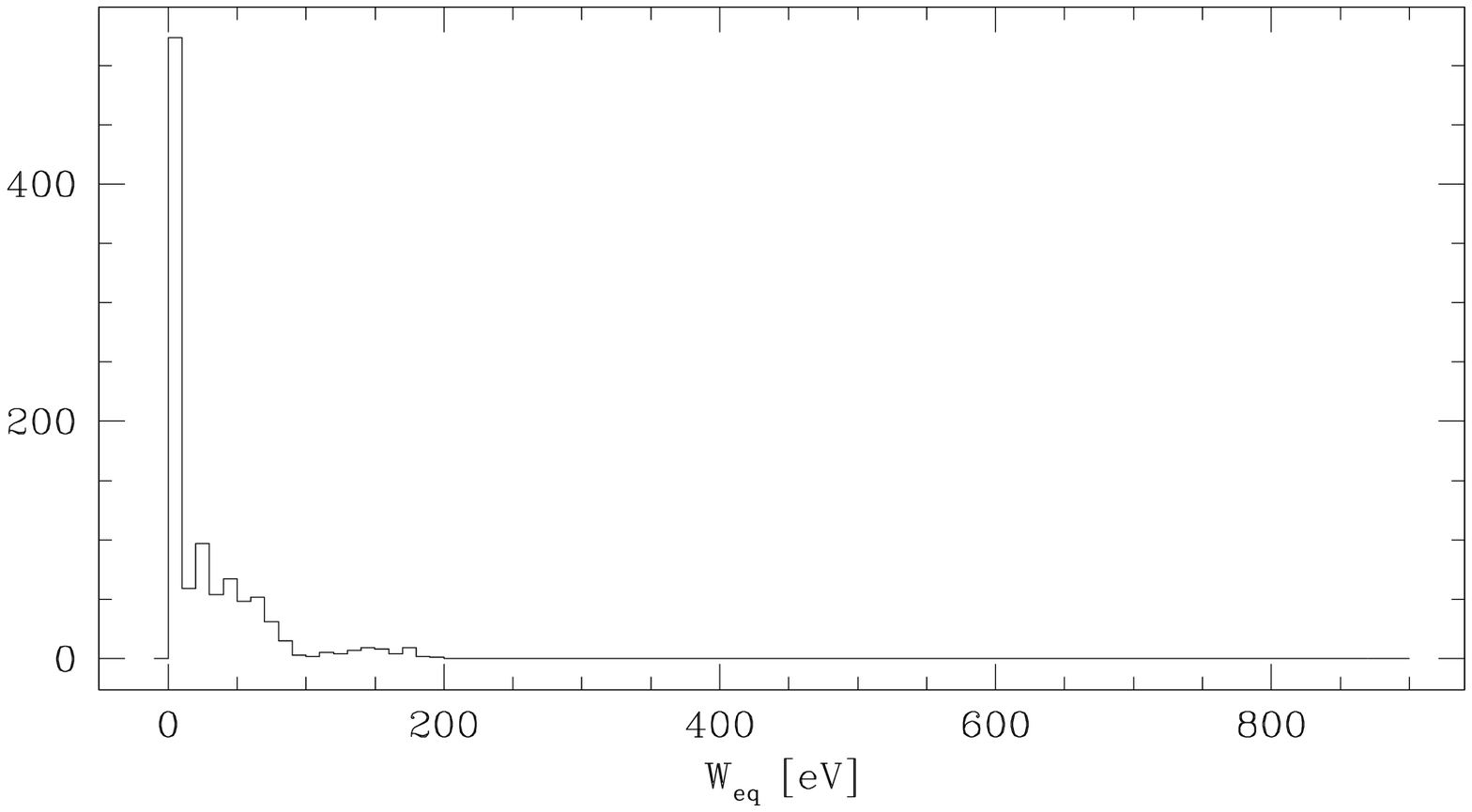, width=9.0cm}
        \epsfig{file=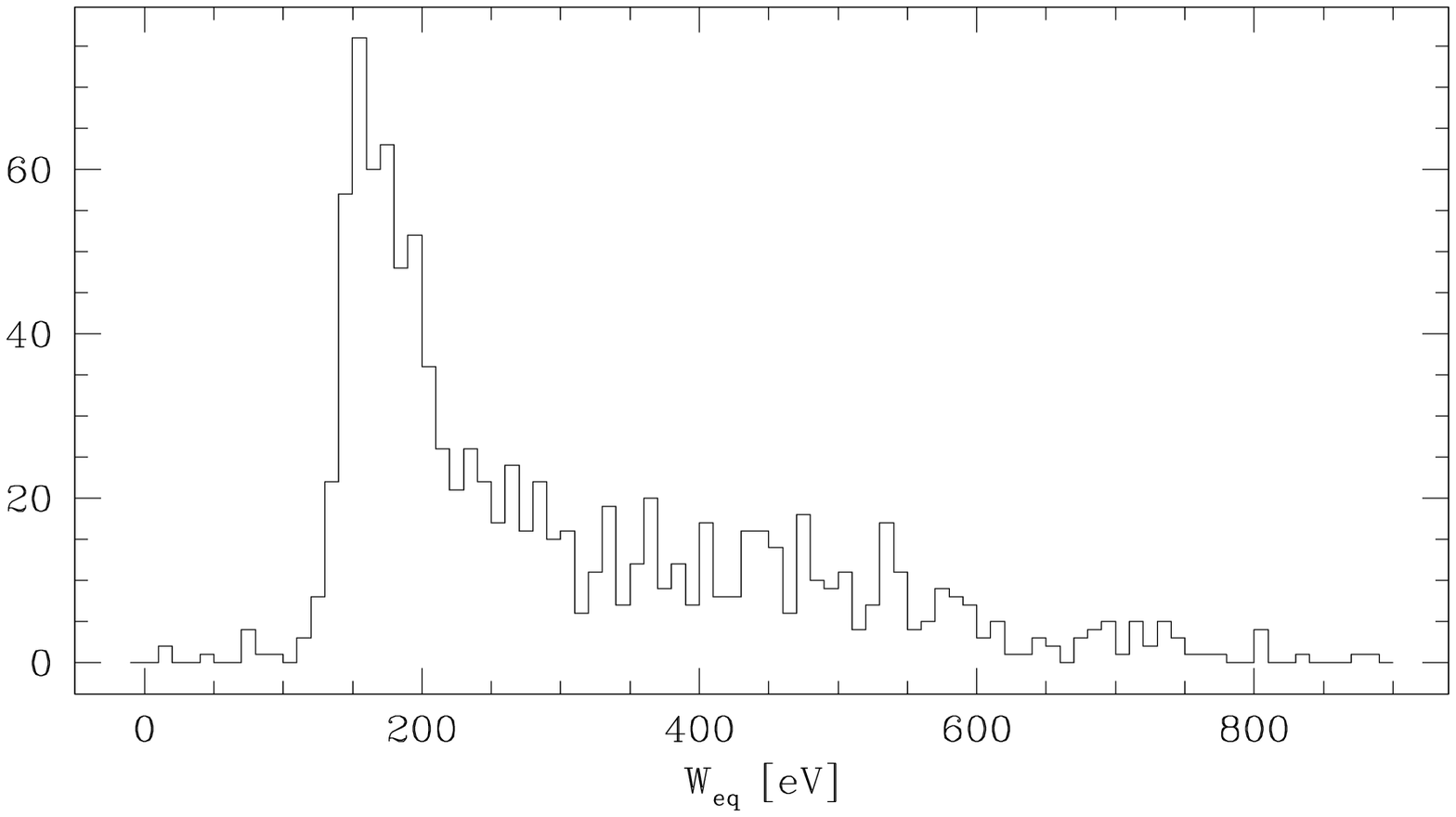, width=9.0cm}

\caption{Distributions of the best-fit equivalent width of the
6.4 keV line in two sets of simulations of the spectra of ``seg1''.  In the
top panel the input model to the simulation did not contain a Gaussian
line component at 6.4 keV, so the distribution gives the probability
of deriving a spurious emission line with a given equivalent width. The
best-fit equivalent width derived for ``seg1'' of 13 eV is consistent with the
line being spurious.  In the bottom panel the input model contained a
Gaussian line component at 6.4 keV with $W_{\rm 6.4~keV} = 249$~eV, as
derived in the data for ``seg2''. In this case, the probability of
detecting a Gaussian component with $W_{\rm 6.4~keV} \le 13$ eV is very
low (0.1\%). Each simulation consisted of 1000 random realisations of
the input model. }

    \label{fig:sim_eqw}
  \end{center}
\end{figure}

To quantify the reliability of the detection of the 6.4 keV emission
in the different spectra of Elias 29, we performed a set of Monte Carlo
simulations  of the PN spectra. For each time interval discussed
in Sect.\,\ref{sec:res}, 1000 random realisations of the absorbed
1T-model fitted to the spectrum {\em without} the additional Gaussian
component at 6.4 keV (as per Table\,\ref{tab:psfit}) were generated
using {\sc xspec}. These random realisations have the same noise
characteristics of the data and the same energy binning of the real
data (for which channels are combined to a minimum of 20 counts per
energy bin).  The 1000 simulated spectra were then fitted with the
same procedure used for the real data. First a fit with an absorbed 1T
plasma model was performed then a Gaussian line component at 6.4 keV
with a narrow $\sigma$ of 10 eV was added to the model and the
simulated data were re-fitted. The line position and width were kept
fixed during this fit, while its normalisation was left
unconstrained. The equivalent width of the fitted line was then
computed for all 1000 fits for each time interval.

Since the input model to the simulated spectra did not contain any
emission line at 6.4 keV, the number (over 1000) of fitted spectra
with a fitted line with a given $W_{\rm 6.4~keV}$ gives the
probability that a line with that equivalent width could be measured
in the real spectrum because of the noise's random fluctuations. For
example, in the case of ``seg1'' the number of simulated spectra with
$W_{\rm 6.4~keV} \ge 13$ eV ($W_{\rm 6.4~keV} = 13$ eV being the
best-fit equivalent width in the real data) is 453, thus, the
probability of deriving a spurious line with $W_{\rm 6.4~keV} \ge 13$
eV from the spectrum of ``seg1'' is 45\%.  This can be seen by looking
at the top panel of Fig.\,\ref{fig:sim_eqw}, which gives the
distribution of $W_{\rm 6.4~keV}$ for the fitted lines for the
simulation of ``seg1''.  Note that, in the case of ``seg1'', there is no
visible excess of emission at 6.4 keV, so a high incidence of
simulated spectra with $W_{\rm 6.4~keV} \ge 13$ eV is consistent with
the absence of the line in the intrinsic spectrum.

\begin{figure}[!htbp]
  \begin{center} \leavevmode 
        \epsfig{file=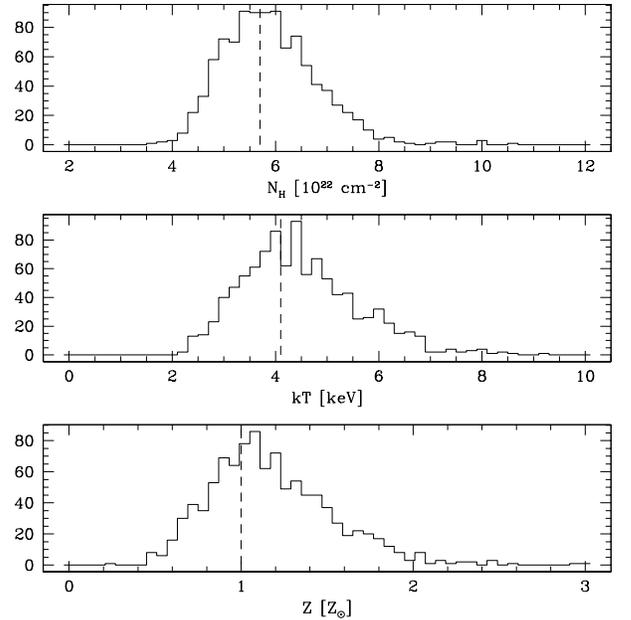, width=9.0cm}

\caption{Distributions of the best-fit values of the absorbing column
density, plasma temperature, and metallicity in the the simulations of
the spectra of ``seg1''. The vertical dashed lines give the parameter
values used in the simulation's input model. The simulation consisted in
1000 random realisations of the spectral model fitted to the data of
``seg1''.}

    \label{fig:sim1_param}
  \end{center}
\end{figure}

In the case of ``seg2'', on the other hand, the number of simulated
spectra with $W_{\rm 6.4~keV} \ge 249$ eV is 1 (over 1000
simulations), indicating that the excess at 6.4 keV seen in the
spectrum of ``seg2'' is very likely intrinsic to the source.  The
results of all these simulations for all the different time intervals,
in terms of the probability of by chance observing a line with a value
of $W_{\rm 6.4~keV}$ greater or equal to the one measured in the data,
are summarised in Table\,\ref{tab:psfit_line}. For illustration, the
distribution of the fitted spectral parameters $N({\rm H})$, $kT$, and
$Z$ for the 1000 simulations for ``seg1'' are given in
Fig.\,\ref{fig:sim1_param}.

Finally, to assess the probability that a line with the same equivalent
width as derived in ``seg2'' ($W_{\rm 6.4~keV} = 249$~eV) could be
present in the source's intrinsic spectrum of ``seg1'', but could go
undetected due to the noise fluctuation, we performed one additional
simulation for ``seg1''. In this case, the simulation input model was
the best-fit absorbed 1T model for ``seg1'' (as per
Table\,\ref{tab:psfit}) with the addition of a Gaussian line at 6.4
keV with $W_{\rm 6.4~keV} = 249$~eV. In this case, the number of
simulated spectra for which we derived a fitted line at 6.4 keV with
$W_{\rm 6.4~keV} \le 13$ eV is 1 (over 1000), as can be seen from
Fig.\,\ref{fig:sim_eqw}. In other words, the probability that during
``seg1'' an intrinsic emission line with the same $W_{\rm 6.4~keV}$ as
derived in ``seg2'' could result in an observed excess with $W_{\rm
6.4~keV} \le 13$ is 0.1\%. This indicates that the strong variations in Fe
6.4 keV emission detected in the data between ``seg1'' and ``seg2''
are very likely intrinsic.

\section{Discussion}
\label{sec:disc}

As mentioned in the introduction, previous works have considered that
the most likely location of the cold material traced by fluorescent
emission is, in YSOs, the circumstellar disk, as no fluorescence,
whether from the photosphere or from a circumstellar shell, can result
in the large equivalent widths observed (approximately 150 eV), given
a plausible thermal photoionising continuum.

The radiative transfer computation necessary for determining the
fluorescent equivalent width in the case of an optically thick target
(i.e.\ a stellar photosphere) shows that, in the case of a
semi-isotropically illuminated slab (such as a photosphere), the
equivalent width $W_{\rm 6.4~keV}$ varies approximately from 150 to 90
eV (assuming photospheric solar iron abundance), for a power-law incident
spectrum with spectral index $\Gamma = 1.2$ to 2.1
(\citealp{gf91}). Given that the thermal spectrum of stellar sources
is generally steeper than a power-law spectrum with $\Gamma \sim 2.0$
(e.g. \citealp{fms+05}; \citealp{gfm+07}), photospheric emission can
hardly explain the observed equivalent widths.

\citet{gf91} show that higher equivalent widths for the fluorescent Fe
K line can be obtained if the cold material is distributed in a
centrally illuminated disk. In this case, for favourable disk
inclination angles (near face-on, $i = 0^{\circ} - 30^{\circ}$),
$W_{\rm 6.4~keV}$ varies approximately between  120 and 200 eV for an
incident spectrum spectral index $\Gamma = 2.3$ to 1.3.  This
interpretation of the origin of the fluorescent Fe K emission  was
also proposed by \citet{fms+05} for Elias 29, whose accretion disk was
shown to be in a relatively face-on orientation ($i < 60^{\circ}$) by
\citet{bhc+02}. This interpretation cannot, however, easily explain the
current set of data. 

The remarkable feature of the present data is that the strong
Fe 6.4 keV emission, apparent in the second segment of the
observation, is absent (or significantly weaker) during the first
segment of the observation, while the observed thermal continuum X-ray
spectrum is basically unchanged. Although a stellar flare takes places
between the two intervals, the radiative lifetimes of fluorescent Fe
transitions are negligible compared to the time between the flare and
the second observation segment (91 ks) -- see e.g. \cite{lsh95} for
the radiative lifetimes of Fe\,{\sc i} and \cite{ssk04} for
Fe\,{\sc ii} -- and therefore the flare cannot explain the 6.4~keV
emission observed in the second segment of the observation. 
To explain the observation, a sustained mechanism ionising the
neutral Fe atoms must operate during the rest of the observation, on
time scale of days, independent of (and not correlated with) the
thermal X-ray emission. This mechanism must be absent during the first
segment. This is unlikely to be the stellar X-ray continuum emission,
as this emission appears to be unchanged between the first and the
second observation segments.

A scenario in which a localised region of photoionised Fe 6.4 keV emission is
somehow eclipsed during ``seg1'' is also difficult to reconcile with
the data, given the high equivalent width of the line ($W_{\rm 6.4~keV}
\ga 150$~eV), when this is present. As mentioned above, to
explain such a large equivalent width by photoionisation, the
favorable geometry of a nearly face-on, centrally illuminated disk
is needed in order to maximise the number of atoms whose line emission
is free to escape. Requiring that the Fe 6.4 keV emitting region is
localised (and in order to be eclipsed by the star in a nearly face-on
disk this volume needs to be significantly localised), automatically
reduces the equivalent widths that can be obtained. The same line of
reasoning applies to a region of localised emission on the stellar
photosphere, even more so, given that for the same incident spectrum
one obtains smaller line equivalent widths.

The large Fe 6.4 keV equivalent width derived in the second segment,
190 -- 250 eV, is also difficult to reconcile with emission from
photoionisation of the material in the circumstellar disk. The model
of \citet{gf91} for a centrally illuminated face-on disk and an
incident X-ray spectrum with a power law index of $\Gamma = 2.3$,
which is appropriate for Elias 29\footnote{as verified by performing a
fit to the ``seg2'' spectrum of Elias 29 with an absorbed power-law
model}, predicts a maximum equivalent width of only 120 eV, that is,
70 -- 130 eV {\em less} than the observed best-fit value in segment
2.  In addition, the maximum equivalent width of the 6.4 keV line
corresponds to the lowest temperature in the thermal spectrum, in
contradiction with the expectations from a photoionisation process.

An alternative explanation for the production of the Fe 6.4 keV
emission line in solar flares is collisional ionisation of K-shell
electrons by a beam of non-thermal electrons (\citealp{epd86}). This
mechanism may also operate in some stellar systems. \citet{odt+07}
have recently observed a superflare on the active binary system
II\,Peg with the \emph{Swift} telescope. Analysis of the X-ray
spectrum from 0.8 to 200 keV from the XRT and BAT instruments reveals
evidence of a thermal component with $kT$ in excess of 7 keV, Fe 6.4
keV emission, and a tail of emission out to 200 keV, which can be fit
with a power-law model with spectral index $\Gamma \sim 2$. They
attribute this power-law hard X-ray emission to non-thermal
thick-target bremsstrahlung emission from a population of accelerated
electrons and find that collisional ionisation (of photospheric
material) from the same non-thermal electrons is a more likely
explanation for the 6.4 keV feature than the photoionisation
mechanism.

In our case, a superflare is not taking place; indeed, no stellar
flare is taking place during the second segment of observation. In a
magnetospheric accretion scenario, however, ionised material is
magnetically channeled in accretion streams that connect the star to
the circumstellar accretion disk (e.g. \citealp{ch92}). The
observation of very intense long-duration flares in YSOs, which
implies flaring loop lengths of tens of stellar radii
(\citealp{ffr+2005}), provides evidence of flaring associated to
these accretion streams.

A magnetic reconnection event leading to the acceleration of electrons
within one of these structures, connecting the star to the disk, may
or may not result in a (soft) X-ray flare event depending on the
density of the material confined in the structure itself.  If the
density of material within the magnetic flux tube is not very high
(with column densities from the acceleration site to the photosphere
or to the disk of $N \la 10^{20}$ cm$^{-2}$), the accelerated
particles will impact the stellar photosphere and heat and
``evaporate'' the chromospheric plasma, providing the hot material
that emits the X-ray flare observed between 0.1 and 10 keV. Indeed,
the flares for which long flaring loops were derived in
\citet{ffr+2005} appear to take place in CTTs with no significant
ongoing accretion (hence their ``accretion'' tubes are not
mass-loaded).

\begin{figure}
  \begin{center} \leavevmode 
        \epsfig{file=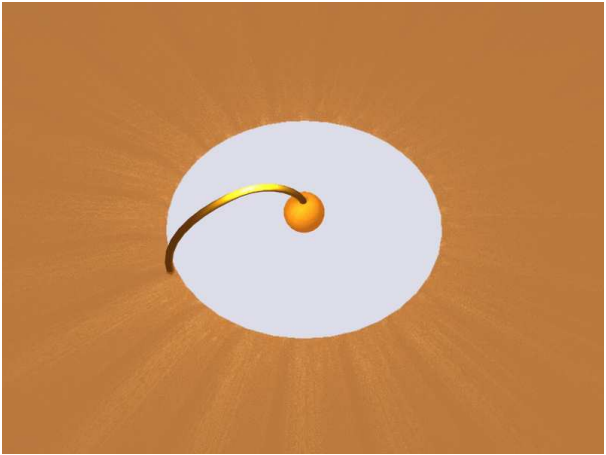, width=8.0cm}
\caption{A cartoon of Elias 29 star-disk system with an accretion tube
(hypothetical). A magnetic reconnection event within the accretion
tube would generate accelerated electrons that could ionise the
material and induce Fe K$\alpha$ emission from the accreting material
within the tube itself. The system is assumed at a viewing inclination
angle $i=30^{\circ}$ ($i=0^{\circ}$ being face-on).}  \label{fig:el29}
\end{center}
\end{figure}

In YSOs, which are still strongly accreting, the density of material
confined in the accretion tubes is expected to be high. For a fiducial
case of mass accretion rate of $10^{-7} M_{\odot}$ yr$^{-1}$,
\citet{mch01} estimate densities of $n_{\rm H} = 10^{12} - 10^{13}$
cm$^{-3}$ at $\sim 2 {\rm R_{*}}$, that is, column densities of the
order of $N \sim 10^{23} - 10^{24}$ cm$^{-2}$ assuming a stellar
radius of 10$^{11}$ cm$^{-2}$. In Elias 29 with an accretion
luminosity of 28.8~$L_{\sun}$, the accretion rate is estimated to be
one order of magnitude higher (\citealp{nts06}), so the column density
within the accretion tubes will be correspondingly higher. With such a
high density, the accelareted electrons produced by a reconnection
event in the accretion tube cannot reach the stellar photosphere, but
will be decelerated in situ, producing non-thermal (hard) X-ray
emission and collisional ionisation of elements such as Fe, which will
be largely neutral at a typical temperature of $\sim 7000$ K
(\citealp{mch01}). A sketch of the Elias 29 star-disk system with the
hypothesised accretion tube where the electron acceleration is taking
place is shown in Fig.\,\ref{fig:el29}.

To explain the observed long-lasting fluorescent emission, the
electron acceleration mechanism should be sustained over several
days. While we have no detailed physical mechanism to propose for
this, we note that many of the long-lasting flares observed in YSOs by
\citet{ffr+2005} require sustained heating of the flaring material to
explain the observed duration of the events, which in some cases
lasted a few days. The thermodynamic cooling time of the flaring
plasma would be much shorter; therefore, to explain the observed
long flaring events in YSOs, magnetic reconnection must be ongoing, on
time scales of days, similar to what is observed for the Fe 6.4 keV
fluorescent emission from Elias 29. The same electron acceleration
mechanism could therefore result in long-lasting flares, if it takes
place in systems with little ongoing accretion, or in strong Fe 6.4
keV emission, if it takes place in systems with strong ongoing
accretion.

In their study, \citet{epd86} provide a relation (Fig. 1 in their
paper) between the flux emitted in the 6.4 keV line in the case of
electron ionisation, as a function of the column density of material
between the electrons' injection point and the ``cold'' Fe (where Fe
6.4 keV emission can take place). This column density is zero in our
case, since, as mentioned above, the material in the accretion tubes
is mostly neutral at temperatures of $\sim 7000$ K.  In the diagram,
different curves are given for different values of the spectral index,
$\Gamma$, of the non-thermal X-ray emission, and the line flux is
normalised to an amplitude of non-thermal (power-law) emission of $a =
1$ photons cm$^{-2}$ s$^{-1}$ keV$^{-1}$ (at 1 keV).

The flux observed for Elias 29 in the Fe 6.4 keV emission line in the
integrated spectrum is $\sim 10^{-6}$ photons cm$^{-2}$ s$^{-1}$, and
from the plot by \citealp{epd86} one can see that such a flux can be
obtained for a very wide choice of parameter combinations, with
$\Gamma$ ranging from 2.1 to 5.0, for $a = 10^{-4} - 1$ photons
cm$^{-2}$ s$^{-1}$ keV$^{-1}$.  The parameters $\Gamma$ and $a$
refer to the hard X-ray emission, typically at $E \ga 20$ keV. As
discussed by \cite{holman03}, however, the X-ray spectrum of the
non-thermal X-ray emission flattens below the low-energy cutoff
($E_c$) of the suprathermal electron responsible for the emission. For
example, the bremsstrahlung emission from a beam of electrons, with
a power-law spectrum with spectral index $\delta = 3$ and low-energy
cutoff $E_c = 50$ keV, has a spectral index of $\Gamma \sim 2$ at $E >
E_c$, but flattens to $\Gamma \sim 1.3$ between 1 and 10 keV. 

For the electron beam model in solar flares, electrons are assumed to
have energy higher than 20 keV (e.g. \citealp{bkm+90}) and recent
RHESSI data indicates a low-energy cutoff that is typically $E_c = 20
- 40$ keV and can be as high 70 keV (\citealp{hss+03}). Using the
BREMTHICK\footnote{available at {\em
http://hesperia.gsfc.nasa.gov/hessi/modelware.htm}} code by
G. D. Holman, we verified that a non thermal hard X-ray emission with
$a = 10^{-4} - 1$ photons cm$^{-2}$ s$^{-1}$ keV$^{-1}$ and $\Gamma =
2-5$, from a population of electrons with $E_{c} \ga 40$ keV, between
1 and 10 keV, would be buried in the star's thermal emission  and thus
 cannot be constrained without hard-X ray observations.  Note, also,
that a value of $a = 10^{-4} - 1$ photons cm$^{-2}$ s$^{-1}$
keV$^{-1}$ would not require an unusually high power input in
accelerated electrons.

If our proposed mechanism of in situ electron bombardment of neutral
or nearly-neutral material is correct, our observations provide
additional evidence of strong non-thermal phenomena in young stars
and, in particular, of particle bombardment of neutral
material in the protoplanetary disk of the young star. In the solar
system, such evidence is, for example, provided by the meteorites known
as chondrites. Chondrites have a number of peculiar features, as they
contain granules (the chondrules) that have been flash heated (on
time scales of less than one hour) to some 2000 K, embedded in a rocky
matrix with no evidence of heating. Also, the chondrules show evidence
of having cooled in a relatively intense magnetic field (of the order
of 10 G). Finally, they contain material with peculiar isotopic
ratios, with significant amounts of short-lived radio nuclides
(e.g. $^{41}$Ca, with a half life of only 0.1 Myr, or $^{10}$Be, with
a half life of 1.5 Myr), which therefore cannot be originating in SN
nucleosynthesis but must instead be originating in situ (as discussed in
detail by e.g.\,\citealp{gss+06}). While the observed isotopic
anomalies likely require proton (rather than electron) bombardment,
our observations provide evidence of in situ
acceleration mechanisms closely associated with neutral material from
the star's circumstellar disk, that is, an environment where
accelerated non-thermal particles interact with the protoplanetary
material from which the meteorites will later grow.

\section{Conclusions}

So far, the Fe 6.4 keV emission in YSOs has been explained in terms of
fluorescent emission from the photoionised (colder) material in the
circumstellar disk. This scenario, however, cannot easily explain the
observed variability of the Fe 6.4 keV emission in Elias 29, which
occurs in the absence of signficant variations of the observed X-ray
continuum. The equivalent width of the line at its maximum strength of
$\sim 250$ eV is also not easily reconciled with a photoionisation
scenario. An alternative line-formation mechanism is collisional
excitation by a population of non-thermal electrons. We suggest that
these electrons could be accelerated by magnetic reconnection events
in the accretion tubes that connect the star to its circumstellar
disk. The electrons are decelerated in situ by the accreting material,
ionising it and causing the observed Fe 6.4 keV emission.

\begin{acknowledgements}

We would like to thank L. Testi and G. Peres for the useful
discussions and an anonymous referee for very helpful comments. EF, GM,
IP, and SS acknowledge the financial contribution from contract ASI-INAF
I/023/05/0.

\end{acknowledgements}

\end{document}